\pdfoutput=1
\documentclass[aps,prl,groupedaddress,showpacs,showkeys,twocolumn,amsmath,
amsfonts,amssymb,superscriptaddress]{revtex4-1}
\usepackage{graphicx}
\usepackage[utf8x]{inputenc}
\usepackage{hyperref}
\usepackage{xcolor}
\usepackage{qcircuit}
\usepackage{physics}
\usepackage{amsbsy}

\usepackage{dcolumn}
\usepackage{bm}

\usepackage[T1]{fontenc}
\usepackage{mathptmx}
\usepackage{etoolbox}
\usepackage{amsfonts}
\usepackage{amsmath}
\usepackage{physics}
\usepackage{enumerate}
\usepackage{enumitem}
\usepackage[normalem]{ulem}
\usepackage{textcomp}
\hypersetup{
    colorlinks,
    linkcolor={red!50!black},
    citecolor={blue!60!black},
    urlcolor={blue!50!black}
}

\begin{document}


\title{Unveiling the geometric meaning of quantum entanglement: discrete and continuous variable systems}
\author{Arthur Vesperini}
\affiliation{DSFTA, University of Siena, Via Roma 56, 53100 Siena, Italy}
\affiliation{QSTAR \& CNR - Istituto Nazionale di Ottica,    Largo Enrico Fermi 2, I-50125 Firenze, Italy}
\affiliation{INFN Sezione di Perugia, I-06123 Perugia, Italy}

\author{Ghofrane Bel-Hadj-Aissa}
\affiliation{DSFTA, University of Siena, Via Roma 56, 53100 Siena, Italy}
\affiliation{QSTAR \& CNR - Istituto Nazionale di Ottica,    Largo Enrico Fermi 2, I-50125 Firenze, Italy}
\affiliation{INFN Sezione di Perugia, I-06123 Perugia, Italy}

\author{Lorenzo Capra}
\affiliation{DSFTA, University of Siena, Via Roma 56, 53100 Siena, Italy}

\author{Roberto Franzosi}
\email[]{roberto.franzosi@unisi.it}
\affiliation{DSFTA, University of Siena, Via Roma 56, 53100 Siena, Italy}
\affiliation{QSTAR \& CNR - Istituto Nazionale di Ottica,    Largo Enrico Fermi 2, I-50125 Firenze, Italy}
\affiliation{INFN Sezione di Perugia, I-06123 Perugia, Italy}

\date{\today}

\begin{abstract}

We show that the manifold of quantum states is endowed with a rich and nontrivial geometric structure. We derive the Fubini-Study metric of the projective Hilbert space of a multi-qubit quantum system, endowing it with a Riemannian metric structure, and investigate its deep link with the entanglement of the states of this space. As a measure, we adopt the \emph{entanglement distance} $E$ preliminary proposed in Ref. \cite{PhysRevA.101.042129}. Our analysis shows that entanglement has a geometric interpretation: $E(|\psi\rangle)$ is the minimum value of the sum of the squared distances between $|\psi\rangle$ and its conjugate states, namely the states ${\bf v}^\mu \cdot {\bm \sigma}^\mu |\psi\rangle$, where ${\bf v}^\mu$ are unit vectors and $\mu$ runs on the number of parties. 
Within the proposed geometric approach, we derive a general method to determine when two states are not the same state up to the action of local unitary operators.

Furthermore, we prove that the entanglement distance, along with its convex roof expansion to mixed states, fulfils the three conditions required for an entanglement measure: that is {\it i)} $E(|\psi\rangle) =0$ iff $|\psi\rangle$ is fully separable; {\it ii)}  $E$ is invariant under local unitary transformations; {\it iii)} $E$  doesn't increase under local operation and classical communications. Two different proofs are provided for this latter property.

We also show that in the case of two qubits pure states, the entanglement distance for a state $|\psi\rangle$ coincides with two times the square of the concurrence of this state.

We propose a generalization of  the entanglement distance to continuous variable systems.

Finally, we apply the proposed geometric approach to the study of the entanglement magnitude and the equivalence classes properties, of three families of states linked to the Greenberger-Horne-Zeilinger states, the Briegel Raussendorf states and the W states.
As an example of application for the case of a system with continuous variables, we have considered a system of two coupled Glauber coherent states.
\end{abstract}


\maketitle



\section{Introduction}
Entanglement is essential in quantum information theory and for its application to quantum technologies. 
Indeed, entanglement is a fundamental resource in quantum cryptography,  teleportation, quantum computation and quantum metrology applications \cite{GUHNE20091,alireza,PhysRevA.105.052439}.
However, entanglement remains elusive since its characterization and quantification in the case of a general system remains an open problem \citep{PhysRevA.95.062116, PhysRevA.67.022320,ScienticReports1-13}.
A huge literature has been devoted to the entanglement quantification problem in the last decades, and despite that, rigorous achievements pertain to bipartite systems case \cite{RevModPhys.81.865}.
In particular, the entropy of entanglement is accepted as a measure for pure states of bipartite systems \cite{PhysRevA.56.R3319}, the entanglement of formation \cite{PhysRevLett.80.2245}, the entanglement distillation \cite{PhysRevA.54.3824, PhysRevLett.76.722, PhysRevLett.80.5239} and related entropies of entanglement \cite{PhysRevLett.78.2275} are acknowledged as faithful measures still for bipartite mixed systems \cite{Adesso_2016}.
A broad literature is devoted to the study of entanglement in multipartite systems. Several approaches have been proposed, 
such that the study of equivalence classes in the case of multipartite entangled pure states
\cite{PhysRevA.62.062314,briegel_PRL86_910},
and the characterization of entanglement by the Schmidt measure or by a generalization of concurrence in the case of mixed multipartite entangled states \cite{PhysRevA.64.022306,PhysRevResearch.2.043062}
or with a generalisation of concurrence, \cite{PhysRevA.61.052306, PhysRevLett.93.230501}.
Also, entanglement estimation-oriented approaches derived from a statistical distance \cite{PhysRevLett.72.3439,EPJD} concept, as the quantum Fisher information \cite{PhysRevLett.102.100401,PhysRevA.85.022321,j.aop.2019.167995}, have been proposed.

In the present work, we derive the Fubini-Study metric \cite{cmp/1103908308, gibbons,  BRODY200119} , which imparts a Riemannian metric structure to the manifold of multi-qubit states. Consequently, we explore the profound connection between the Riemannian metric structure associated with the projective Hilbert space of a quantum system and the entanglement of the states within this space.  The entanglement measure that we have adopted for our analysis is the entanglement distance (ED) $E$, a measure preliminary proposed in Ref. \cite{PhysRevA.101.042129} by some of us. Our investigation shows that entanglement has a geometric interpretation. In fact, $E(|\psi\rangle)$ is the minimum value of the sum of the squared distances between $|\psi\rangle$ and its conjugate states, namely the states ${\bf v}^\mu \cdot {\bm \sigma}^\mu |\psi\rangle$, where ${\bf v}^\mu$ are unit vectors and $\mu$ runs on the number of parties. 
Furthermore, the proposed geometric approach, allows us to derive a general method to determine if, or not, two states, actually are the same state up to the action of local unitary (LU) operators.

Furthermore, we show that the ED, along with its convex roof expansion to mixed states, is an entanglement monotone in the sense of \cite{vidal_2000,PhysRevLett.78.2275} that is, it fulfils the three following conditions: {\it i)} $E(|\psi\rangle) =0$ iff $|\psi\rangle$ is fully separable; {\it ii)}  $E$ is invariant under LU transformation; {\it iii)} $E$  doesn't increase, on average, under local operation and classical communications (LOCC). Two different proofs are provided for this latter property.

We also show that in the case of a two qubits pure state, the entanglement distance for a state $|\psi\rangle$ coincides with twice the square of the concurrence of this state.

We then propose a necessary condition (sufficient for $M=2$) for the LU equivalence of two pure quantum states, relying on the local properties of the associated Fubini-Study metrics.

In addition, we propose an extension of the entanglement distance to systems with continuous variables, and we consider its application to a system described by products of coherent Glauber states.

Finally, we report some examples of the application of the proposed geometric approach to three parameter-dependent families of states, derived from the Greenberger-Horne-Zeilinger states \cite{ghz}, the Briegel Raussendorf states \cite{briegel_PRL86_910} and the W states \cite{PhysRevA.62.062314}.
We have shown that for $M=2$ the three families belong to the same class. For $M=3$ the family of Greenberger-Horne-Zeilinger states and the family of Briegel Raussendorf states, belong to the same class, whereas for $M=4$ the three families belong to disjoint classes.

\section{Geometry of projective Hilbert space}
Quantum mechanics is essentially a geometric theory. In that sense, a useful geometrical tool is that of the Riemannian metric structure associated with the manifold of states of quantum mechanics. The Hilbert space is endowed with a Hermitian scalar product that naturally induces a distance between vectors. If ${\cal H}$ denotes a Hilbert space of a general quantum system, for given two close vectors in ${\cal H}$, $| \psi_1 \rangle$ and $ |\psi_2\rangle$, from the scalar product $\langle \psi_1|\psi_2\rangle$, one derives the norm $\Vert \Vert$ and the (finite) distance between these two vectors as
\begin{equation}
D(| \psi_1 \rangle,|\psi_2\rangle)=\Vert | \psi_1 \rangle - |\psi_2\rangle\Vert =
\langle \psi| \psi \rangle^{1/2} \, ,
\end{equation}
where $|\psi\rangle = |\psi_1\rangle -|\psi_2\rangle$. In the case of 
two normalized vectors $| \psi_1 \rangle$ and $ |\psi_2\rangle$, it results
\begin{equation}
D(| \psi_1 \rangle,|\psi_2\rangle)=  \left[ 2 \left( 1- \Re (\langle \psi_1  |\psi_2\rangle ) \right) \right]^{1/2} \, .
\end{equation}
Furthermore, each Hilbert space has a structure of a differentiable manifold, so it is always possible to define a local chart on ${\cal H}$, which includes two close states.  This allows one to derive the metric tensor induced by the above-defined distance. 
Let $|\psi \rangle$ and $|\psi\rangle + |d\psi\rangle$, two close vectors. The squared (differential) distance between these is derived by developing up to the second order $D$, and it results
\begin{equation}
d^2(|\psi\rangle + |d\psi\rangle,|\psi\rangle)=\langle d\psi  |d\psi\rangle \, .
\end{equation}
Thus, by means of a local chart, the normalized vectors in ${\cal H}$ smoothly depend on $N$-dimensional parameter $\xi\in \mathbb{R}^N$ and one has
\begin{equation}
|d\psi\rangle = \sum_\mu |\partial_\mu \psi(\xi) \rangle d\xi^\mu \, ,
\end{equation}
where with $\partial_\mu \psi$ we mean ${\partial \psi}/{\partial \xi^\mu}$. 
Thus one has
\begin{equation}
d^2(|\psi\rangle + |d\psi\rangle,|\psi\rangle) = \sum_{\mu \nu}\langle \partial_\mu \psi|\partial_\nu \psi \rangle d\xi^\nu  d\xi^\mu \, .
\end{equation}

Despite the matrix elements $\langle\partial_\mu \psi | \partial_\nu \psi \rangle$ might seem the entries of a Riemannian metric tensor for ${\cal H}$, they do not have any physical interpretation for the distance  between states.
In fact, the Hilbert space provides a redundant description of quantum states. These latter are associated with the rays of the Hilbert space, and two normalized kets differing by a phase factor $e^{i\alpha}$, represent the same quantum state. Consistently, the distance between $|\psi\rangle$ and $|\psi\rangle + |d\psi\rangle$, and the one between $|\psi^\prime\rangle =e^{i\alpha}|\psi\rangle$ and $|\psi^\prime\rangle + |d \psi^\prime\rangle$ must be the same.  By resorting to a local chart, we  may express this request in a mathematical framework. An appropriate metric tensor for the states space has to be invariant under gauge transformation $|\psi(\xi)\rangle \to e^{i\alpha(\xi)} |\psi(\xi)\rangle$.
This is accomplished with the Fubini-Study metric which gives the (squared) distance between two neighbouring rays
\begin{equation}
d^2_{FS}(| \psi\rangle + |d \psi\rangle,|\psi\rangle)=\langle d\psi |d\psi\rangle -\langle \psi |d\psi\rangle \langle d\psi|\psi\rangle \, ,
\label{F-S-metric}
\end{equation}
from which one derives the metric tensor
\begin{equation}
g_{\mu \nu} = \langle \partial_\mu \psi|\partial_\nu \psi \rangle -
\langle \partial_\mu \psi| \psi \rangle \langle \psi|\partial_\nu \psi \rangle \, .
\label{mtFS}
\end{equation}
The  Fubini-Study metric \eqref{F-S-metric} is therefore defined on the finite projective Hilbert-space ${\cal PH}$ \cite{cmp/1103908308,gibbons}, that is on the set of equivalence classes of non-zero vectors $|\psi\rangle \in {\cal H}$, for the relation $\thicksim_p $ on ${\cal H}$ given by  $|\psi\rangle  \thicksim_p |\phi\rangle$ iff $|\psi\rangle = \alpha |\phi\rangle$, for some $\alpha \in \mathbb{C}$, $\alpha \neq 0$.

It is worth remarking that one can define the square of the (finite) distance between two rays $[|\phi_1\rangle]_p,[|\phi_2\rangle]_p \in {\cal PH}$, associated with the normalized states $e^{i\alpha_1}|\phi_1\rangle,e^{i\alpha_2}|\phi_2\rangle$, respectively, as follows
\begin{equation}
D^2_{FS}(|\phi_1\rangle,|\phi_2\rangle) = (1-|\langle\phi_1|\phi_2 \rangle|^2) \, .
\label{fFS}
\end{equation}
One can easily verify that the latter distance induces the metric tensor \eqref{mtFS}.
In fact, by expanding $|\phi_1\rangle$ up to second order as 
\begin{equation}
|\phi_1(\xi)\rangle=
|\psi\rangle + \sum_\mu |\partial_\mu \psi \rangle  d\xi^\mu + \frac{1}{2}\sum_{\mu\nu} |\partial^{2}_{\mu\nu} \psi \rangle d\xi^\mu d\xi^\nu \, ,
\end{equation}
and setting $|\phi_2\rangle =|\psi\rangle$, from  Eq. \eqref{fFS} one gets
\begin{equation}
D^2_{FS}(|\phi_1\rangle ,|\phi_2\rangle) =\sum_{\mu\nu} g_{\mu\nu} d\xi^\mu d\xi^\nu \, ,
\label{fFSdiff}
\end{equation}
where $g_{\mu\nu}$ is the one of Eq. \eqref{mtFS}.

\section{\label{sec:pure} The geometric meaning of entanglement measure}
As above mentioned, the present study aims at investigating the deep link between the Riemannian metric structure associated with the projective Hilbert space and the entanglement properties of the states of this space. To this end, we endow the projective Hilbert space with a metric, derived from the Fubini-Study metric, that has the desirable property of making it an attractive definition for entanglement measure.
We consider the case of the Hilbert space
${\cal H } = {\cal H}^0 \otimes {\cal H}^1 \cdots {\cal H}^{M-1}$ tensor product of $M$ qubits Hilbert spaces. 

The entanglement measure is invariant under LU transformations. Thus, 
given $[|\phi\rangle]_p,[|\psi\rangle]_p \in {\cal PH}$ and  the associated normalized vectors $|\phi\rangle,|\psi\rangle \in {\cal H}$, we define the following equivalence relation between elements of the projective Hilbert space
\begin{equation}
[|\phi\rangle]_p \thicksim [|\psi\rangle]_p \, , \quad \rm {iff} \ |\phi\rangle = e^{i\alpha} \prod^{M-1}_{\mu=0} U^\mu |\psi\rangle \, , 
\end{equation}
where, for $\mu=0,\ldots,M-1$, each operator $U^\mu$ is an arbitrary $SU(2)$ LU operator that operates on the $\mu$th qubit and $\alpha \in \mathbb{R}$. With this equivalence relation, one derives the quotient set ${\cal P H}/\thicksim $. Thus, the entanglement measure $E$ has to be a function $E: {\cal PH}/\thicksim \to \mathbb{R}^+$, that is a function of the equivalence classes of ${\cal P H}$ by $\thicksim$, that is
\begin{equation}
[|\psi\rangle] = \left\lbrace |\phi\rangle \in {\cal P H} | \  |\phi\rangle \thicksim |\psi \rangle  \right\rbrace\, .
\label{Us}
\end{equation}
Following Ref. \cite{PhysRevA.101.042129}, we derive an entanglement measure from a distance inspired from the Fubini-Study one. For each normalized ket $|\psi\rangle\in{\cal H}$  we consider
\begin{equation}
\left\lbrace |U,\psi\rangle = \prod^{M-1}_{\mu=0} U^\mu |\psi\rangle \right\rbrace\, ,
\end{equation}
the set of all the vectors derived from $|\psi\rangle$ under the action of LU operators,
where, for $\mu=0,\ldots,M-1$, each operator $U^\mu$ is an arbitrary $SU(2)$ LU operator that operates on the $\mu$th qubit. Note that the set of kets  in \eqref{Us}, derived by varying the operators $U^\mu$, is the class $[|\psi\rangle]$, and also that all these kets have the same degree of entanglement.
For each vector $|U,\psi\rangle $ in \eqref{Us}, we define a local chart in a neighbourhood of it, by means of the unitary operator
$
e^{-i\sum_{\mu=0}^{M-1} \bm{\sigma}_{\bf n}^\mu\xi^\mu } \, ,
$
depending on real parameters $\xi^\mu$, and
where ${\bf n}^\mu$ are assigned unit vectors. In this way, to the point $\xi^\mu =0$, for $\mu=0,\ldots,M-1$, corresponds the vector $|U,\psi\rangle $. 
Here and in the following we use the notation $\bm{\sigma}_{\bf n}^\mu = {\bf n}^\mu \cdot \bm{\sigma}^\mu$, and 
for $\mu=0,\ldots,M-1$, we denote by $\sigma^\mu_1$, $\sigma^\mu_2$ and $\sigma^\mu_3$
the three Pauli matrices operating on the $\mu$-th qubit, where the index $\mu$ labels the spins.
We consider an infinitesimal variation of ket $|U,\psi\rangle $ given by
\begin{equation}
|dU,\psi\rangle = \sum^{M-1}_{\mu=0} d\tilde{U}^\mu |U,\psi\rangle \, ,
\label{dUs}
\end{equation}
where 
\begin{equation}
d\tilde{U}^\mu = -i 
 \bm{\sigma}_{\bf n}^\mu
d \xi^\mu 
\label{dU}
\end{equation}
rotates the $\mu$th qubit by an infinitesimal angle $2 d\xi^\mu$  around
the unitary vector ${\bf n}^\mu $.

By substituting $|U,\psi\rangle$ and $|dU,\psi\rangle $ in Eq. \eqref{F-S-metric}, in place of $|\psi\rangle$ and $|d\psi\rangle$, respectively, we get 
\begin{equation}
    d^2_{_{FS}} (|U,\psi\rangle +|dU,\psi\rangle,|U,\psi\rangle ) =
    \sum_{\mu \nu }g_{\mu \nu} (|\psi\rangle,{\bf v}) d\xi^\mu d\xi^\nu \, ,
\end{equation}
where, the corresponding \textit{projective Fubini-Study metric tensor} is
\begin{equation}
    g_{\mu \nu} (|\psi\rangle,{\bf v}) =
\langle \psi | \bm{\sigma}_{\bf v}^\mu
\bm{\sigma}_{\bf v}^\nu|\psi\rangle 
-\langle \psi | \bm{\sigma}_{\bf v}^\mu
|\psi\rangle
\langle \psi |
\bm{\sigma}_{\bf v}^\nu|\psi\rangle \, ,
\label{gmunu}
\end{equation}
${\bf v} = ({\bf v}^0,\ldots,{\bf v}^{M-1})$ and
the unit vectors ${\bf v}^\mu$, $\mu=0,\dots,M-1$, are derived by a rotation of the original
ones of Eq. \eqref{dU}, according to $\bm{\sigma}^\nu_{\bf v} =
U^{\nu \dagger}  \bm{\sigma}^\nu_{\bf n}
U^{\nu}$,
where there is no summation on the index $\nu$.
Of course, for each state $|\psi\rangle$, the metric tensor $g_{\mu \nu} (|\psi\rangle,{\bf v})$ is not invariant under rotation of the unit vectors ${\bf v}^\mu$. In order to derive a measure that is invariant under rotation of the unit vectors, we define the entanglement measure of $[|\psi\rangle]$, as the inferior value of the trace of $g_{\mu \nu} (|\psi\rangle,{\bf v})$ over all the possible orientations of the unit vectors ${\bf v}^\mu$. In formulas we define the ED as
\begin{equation}
E(|\psi\rangle) =\inf_{ \{{\bf v}^\nu\}_\nu} \tr (g(|\psi\rangle,{\bf v})) \, ,
\label{emeasure}
\end{equation}
where $\tr$ is the trace operator and where the $\inf$ is taken over all possible orientations of the unit vectors $ {\bf v}^\nu$ ($\nu = 0, \dots , M-1$).
We emphasize that, in general, the inspection of the block structure of $g(|\psi\rangle)$ is informative about $k$-separability. Consider a choice of unit vectors $ {\bf v}^\nu$, giving rise to a metric $g(|\psi\rangle,{\bf v})$ which is, up to permutation of the qubits' indices, diagonal by blocks. 
In a previous paper from one of the authors \cite{VESPERINI2023169406}, it was shown that $n\geq p\geq k$, with $n$ the number of such blocks, $p$ the \textit{persistency of entanglement} and $k$ the \textit{degree of separability}.
In particular, this implies that if there exists a given choice of unit vectors yielding $g(|\psi\rangle,{\bf v})$ irreducible for any permutation of its indices (i.e. $n=1$), then $|\psi\rangle$ is genuinely multipartite entangled (i.e. $k=1$).

From Eq. \eqref{gmunu} we derive
\begin{equation}
    \tr [ g(|\psi\rangle,{\bf v}) ]= \sum^{M-1}_{\mu=0} \left[1 - ({\bf v}^\mu\cdot \langle \psi |\bm{\sigma}^\mu|\psi\rangle )^2\right] \, ,
\label{ed1}
\end{equation}
that shows that the unit vectors
\begin{equation}
\tilde{\bf v}^\mu = \pm \langle \psi |\bm{\sigma}^\mu|\psi\rangle/\Vert \langle \psi |\bm{\sigma}^\mu|\psi\rangle \Vert \, ,
    \label{vtildes}
\end{equation}
provide the $\inf$ of $\tr(g)$.
Therefore, we obtain the following directly computable formula for the ED
\begin{equation}
    E(|\psi\rangle) =M - \sum^{M-1}_{\mu =0} \Vert \langle \psi |\bm{\sigma}^\mu|\psi\rangle\Vert^2 \, .
    \label{measureMqpure}
\end{equation}
Note that the latter equation can be seen as the sum of the $M$ single-qubit EDs \begin{equation}\label{measureMqpure_1qubit}
E_\mu(|\psi\rangle)=1 - \Vert \langle \psi |\bm{\sigma}^\mu|\psi\rangle\Vert^2.
\end{equation}
$E_\mu(|\psi\rangle)$ is a measure of bipartite entanglement of $\mu$ with the rest of the system. Note that Eq. \eqref{measureMqpure} also has the meaning of a quantum correlation measure \cite{ScienticReports1-13}.

The $\inf$ operation, makes the measure \eqref{emeasure}
independent from the choice of the operators $U^\mu$. Consequently, its numerical value is associated with the class \eqref{Us}, and does not depend on a specific element chosen inside the class. This is a necessary condition for a well defined entanglement measure \cite{PhysRevLett.78.2275}.

The entanglement measure can be derived by a minimum distance principle, if studied in the framework of the Riemannian geometry of the projective Hilbert space.
In fact, according to Eq. \eqref{fFS} the square of the distance between the rays associated with the unit vectors $|\phi\rangle$ and $|\phi^\mu({\bf v}^\mu)\rangle\equiv\bm{\sigma}_{\bf v}^\mu |\phi\rangle$, is
\begin{equation}
D^2_{FS}(|\phi\rangle,|\phi^\mu({\bf v}^\mu)\rangle) =
1-|\langle \phi | \phi^\mu({\bf v}^\mu)\rangle|^2 \, .
\end{equation}
We name  ${\bf v}^\mu$-conjugate of $|\phi\rangle$ the states $|\phi^\mu({\bf v}^\mu)\rangle$, for $\mu=0,\ldots M-1$.
Therefore 
\begin{equation}
E(|\phi\rangle) = \inf_{ \{{\bf v}^\nu\}_\nu} \sum^{M-1}_{\mu=0}
D^2_{FS}(|\phi\rangle,|\phi^\mu({\bf v}^\mu)\rangle) \, .
\end{equation}
This shows that the minimum of the sum of the semi-square of the (finite) distances between a state $|\phi\rangle$ and all the states derived under the action of the operators $\bm{\sigma}_{\bf v}^\mu$, obtained by varying the vectors ${\bf v}^\mu$, is bounded from below by the entanglement measure $E(|\phi\rangle)$.
For fully separable states, the minimum distance is zero whereas, for maximally entangled states, it is $M$ at the very best.
Therefore, in this geometric framework, entanglement represents an obstacle to the minimum of the sum of the distance square between a state $|\phi\rangle$ and all its ${\bf v}^\mu$-conjugate states.

\section{Geometric characterization of the equivalence classes}
One of the basic questions that this geometric approach can answer, is to determine when two states certainly do not belong to the same equivalence class. Two states sharing the same degree of entanglement might indeed be LU equivalent or not.
Via the study of the full metric tensor associated with two given states, with the same entanglement magnitude, one can determine if these two states belong to different equivalence classes.

Let us consider an equivalence class, let it say $[|\phi\rangle]$, and let us consider two states $|\phi_1\rangle,|\phi_2\rangle \in [|\phi\rangle]$. Then, there exist $M$ LU operators $U^\mu$,
$\mu=0,\ldots,M-1$, each one operating on the $\mu$th qubit, and $\alpha\in\mathbb{R}$ such that
\begin{equation}
|\phi_1 \rangle = e^{i\alpha} U |\phi_2\rangle \, ,
\end{equation}
where $U=\prod^{M-1}_{\mu=0} U^\mu$
We can write
\begin{equation}
\begin{split}
&\langle \phi_1 | \bm{\sigma}_{{\bf v}}^\mu
\bm{\sigma}_{{\bf v}}^\nu|\phi_1\rangle 
-\langle \phi_1 | \bm{\sigma}_{{\bf v}}^\mu
|\phi_1\rangle
\langle \phi_1 |
\bm{\sigma}_{{\bf v}}^\nu|\phi_1\rangle \\ =&
\langle \phi_2 |U^\dagger \bm{\sigma}_{{\bf v}}^\mu
\bm{\sigma}_{{\bf v}}^\nu U|\phi_2\rangle 
-\langle \phi_2 |U^\dagger \bm{\sigma}_{{\bf v}}^\mu
U|\phi_2\rangle
\langle \phi_2 | U^\dagger
\bm{\sigma}_{{\bf v}}^\nu U|\phi_2\rangle \\ =&
\langle \phi_2 | \bm{\sigma}_{{\bf n}}^\mu
\bm{\sigma}_{{\bf n}}^\nu|\phi_2\rangle 
-\langle \phi_2 | \bm{\sigma}_{{\bf n}}^\mu
|\phi_2\rangle
\langle \phi_2 |
\bm{\sigma}_{{\bf n}}^\nu|\phi_2\rangle\, ,
\end{split}
\end{equation}
where $\bm{\sigma}_{{\bf n}}^\mu=
U^{\mu \dagger} \bm{\sigma}_{{\bf v}}^\mu
U^{\mu}$ for $\mu=0,\ldots,M-1$.
Thus, it results
\begin{equation}
g_{\mu \nu} (|\phi_1\rangle,{\bf v}) 
    = 
g_{\mu \nu} (|\phi_2\rangle,{\bf n})
\, .
\label{gmunuclass}
\end{equation}

Therefore, two states do not belong to the same equivalence class if there exists at least a set of unit vectors ${\bf v}^\mu$, $\mu = 0,\ldots,M-1$, such that it is not possible to determine a set of unit vectors ${\bf n}^\mu$, $\mu = 0,\ldots,M-1$, satisfying Eq. \eqref{gmunuclass} for $\mu,\nu=0,\ldots, M-1$. Formally,
\begin{gather}
|\phi_1\rangle \underset{LU}{\sim} |\phi_2\rangle 
\implies 
\forall\{{\bf v}^\mu\}_\mu, \exists\{{\bf n}^\mu\}_\mu\;\big|\; g(|\phi_1\rangle,{\bf v})=g (|\phi_2\rangle,{\bf n}).
\label{eqv-formal}
\end{gather}

Note that, in the general case, one can't use Eq. \eqref{eqv-formal} to determine with certainty if two states belong to the same equivalence class, as the EM encodes only informations up until the second order of the associated quantum statistical distributions, namely two-qubits correlators. In other words, it is possible for Eq. \eqref{gmunuclass} to holds for any choice of ${\bf n}^\mu$, while $|\phi_1\rangle$ and $|\phi_2\rangle$ are not equivalent, in some higher order statistical property.

Clearly this is not the case if $M=2$, since the full statistics do not possess any higher order property. For $M=2$, the implication \eqref{eqv-formal} is both ways, hence the second member stands as a sufficient and necessary condition for state equivalence.

By using the unit vectors $\tilde{\bf v}^\nu$ of Eq. \eqref{vtildes} in place of the ${\bf v}^\nu$ in Eq. \eqref{gmunu}, we get the matrix
\begin{equation}
\tilde{g} (|\psi\rangle)  = g(|\psi\rangle,\tilde{\bf v}) \, ,
\label{EM}
\end{equation}
that we name entanglement metric (EM).
Thus, for two states $|\phi_1\rangle$ and $|\phi_2\rangle$ that differ
from one another under the action of LU transformations, we have 
\begin{equation}
\tilde{g} (|\phi_1\rangle)  = g(|\phi_2\rangle,{\bf n}) \, , 
\end{equation}
where the unit vectors ${\bf n}^\nu$ are derived using suitable rotations of the unit vectors $\tilde{\bf v}^\nu$ provided by Eq. \eqref{vtildes} in the case of state $|\phi_2\rangle$.
Note, in general, the unit vectors ${\bf n}^\nu$ do not coincide with the ones given in Eq. \eqref{vtildes} for the state $|\phi_2\rangle$.
In the following, we will apply this geometric approach to the states belonging to three class states to verify if they are in the same equivalence class or not.

\section{Monotonicity of the Entanglement Distance.}\label{sec:monotonicity}

The single-qubit ED \eqref{measureMqpure_1qubit} can straightforwardly be generalized to mixed state via a convex roof construction, i.e.
\begin{equation}\label{mixedstate_ED}
E_\mu(\rho):=\min_{\{p_j,\psi_j\}}\sum_j p_j E_\mu(|\psi_j\rangle),
\end{equation}
where the minimization is carried over all of the possible realizations $\{p_j,\psi_j\}$ of $\rho$ as a mixture of pure states \footnote{Note that the other formula, proposed in Ref. \cite{ScienticReports1-13} as a generalization of the ED to mixed state, in fact reduces to Eq. \eqref{mixedstate_ED}, and is hence also an entanglement monotone. The supplementary minimization process in the former serves only as a trick, which sometimes allow to overcome the difficulty of the usual minimization over all possible realizations $\{p_j,\psi_j\}$ of $\rho$ as mixture of pure states.}.

A measure is called a measure of entanglement or entanglement monotone if it satisfies the following necessary conditions \cite{PhysRevLett.78.2275,vidal_2000}:

\begin{enumerate}[label=(\roman*)]
\item \label{uno} $E(\rho) =0$ iff $\rho$ is fully separable.

\item \label{due} $E(\rho)$ is invariant under LU operations, i.e.,
$E(\rho) = E(\prod^{M-1}_{\mu=0} U^\mu\rho U^{\mu\dagger})$, where
for $\mu=0,\ldots,M-1$, each operator $U^\mu$ is an arbitrary $SU(2)$ LU operator that operates on the $\mu$th qubit.

\item \label{tre}$E(\rho)$ cannot increase, on average, under local operations and classical communication.
\end{enumerate}

If these three conditions are met for a single-qubit measure as \eqref{mixedstate_ED}, they are also met by their sum, the total ED $E(\rho)=\sum_\mu E_\mu(\rho$).

Let us now prove that the Entanglement Distance satisfies these three conditions.

From Eq. \eqref{emeasure} we have $E(|\psi\rangle) =0$ iff $\inf_{{\bf v}^\mu} g_{\mu\mu}(|\psi\rangle, {\bf v}^\mu) =0$, for  $\mu=0,\ldots,M-1$. From \eqref{gmunu} it results 
$\inf_{{\bf v}^\mu} g_{\mu\mu}(|\psi\rangle, {\bf v}^\mu) =0$ iff $\forall \mu$ it exist ${\bf v}^\mu$ such that $\bm{\sigma}_{\bf v}^\mu |\psi\rangle = |\psi\rangle$. 
Therefore $|\psi\rangle$ is simultaneously eigenstate for $\bm{\sigma}_{\bf v}^\mu$ with the maximum eigenvalue ($+1$), this is possible iff $|\psi\rangle$ is product of single-party eigenstates for $\bm{\sigma}_{\bf v}^\mu$, for $\mu=0,\ldots,M-1$. This prove condition \ref{uno}.

To prove condition \ref{due}, we start from Eq. \eqref{ed1}, by considering $U= \prod^{M-1}_{\nu=0}U^\nu$, where each operator $U^\nu$ is an arbitrary $SU(2)$ LU operator that operates on the $\nu$th qubit.
We have
\begin{equation}\label{LUinv_proof}
\begin{aligned}
E(U|\psi\rangle) =
& \inf_{\{ {\bf v}^\mu \}_\mu} \sum^{M-1}_{\mu=0} \left[1 - ( \langle \psi | U^{\mu\dagger }\bm{\sigma}_{\bf v}^\mu U^\mu|\psi\rangle )^2\right] =
\\
& \inf_{\{ {\bf u}^\mu \}_\mu} \sum^{M-1}_{\mu=0} \left[1 - ( \langle \psi |  (\bm{\sigma}_{\bf u})^\mu |\psi\rangle )^2\right] = E(|\psi\rangle)\, , 
\end{aligned}
\end{equation}
where ${\bf u}^\nu \cdot \bm{\sigma}^\nu =
U^{\nu \dagger} {\bf v}^\nu \cdot \bm{\sigma}^\nu
U^{\nu}$ for $\nu=0,\ldots,M-1$. 
Both these properties are inherited by the related measure on mixed state \eqref{mixedstate_ED}.

Two different proofs of condition \ref{tre} can be found in Appendix.

\section{Comparison between the concurrence and the entanglement distance.}

Let us consider a general $M=2$ qubits normalized pure-state
\begin{equation}
    |\psi\rangle = \sum^3_{j=0} w_j |j\rangle \, ,
    \label{2btstate}
\end{equation}
such thate $  \sum^3_{j=0} |w_j|^2 = 1$.
The concurrence for pure state \eqref{2btstate} is defined as \cite{10.5555/2011326.2011329}
\begin{equation}
    C(|\psi\rangle )= |\langle \psi|\psi^\dagger\rangle| \, ,
    \label{concupsi}
\end{equation}
where $    |\psi^\dagger \rangle = \sigma^0_2 \otimes \sigma^1_2  \sum^3_{j=0} w^*_j |j\rangle $.
By direct computations one gets \cite{10.5555/2011326.2011329}
\begin{equation}
    C(|\psi\rangle) = 2 |w_0 w_3-w_1 w_2| \, .
    \label{concupsi1}
\end{equation}
By a direct calculation, from Eq. \eqref{measureMqpure} one derives for the same general state 
\begin{equation}
    E(|\phi\rangle )= 8[w^2_0 w^2_3 + w^2_1 w^2_2-\! w^*_0w^*_3 w_1 w_2-\! w_0w_3 w^*_1 w^*_2]
    \, .
\end{equation}
Therefore we get the following general result for $M=2$ qubits states
\begin{equation}
    E(|\phi\rangle)/2 = [C(|\phi\rangle)]^2 \, .
    \label{EMvsCpure}
\end{equation}
This proves that the concurrence for pure states, is a special case of ED, valid for the case $M=2$.

\section{Continuous variable systems}
Continuous variable systems (CVS) are described by a Hilbert space ${\cal H}^{cv} = \otimes^n_{\mu=1} {\cal H}_\mu$, where each ${\cal H}_\mu$ is an infinite-dimensional Fock space. Let $a_\mu$ and $a^\dagger_\mu$ be the usual annihilation and creation operators acting on ${\cal H}_\mu$, and $\hat{q}_\mu = i (a_\mu - a^\dagger_\mu)/\sqrt{2}$ and $\hat{p}_\mu = (a_\mu + a^\dagger_\mu)/\sqrt{2}$, the related quadrature phase operators. Therefore, it results in $[a_\mu, a^\dagger_\nu] = \delta_{\mu,\nu}$ and $[\hat{q}_\mu, \hat{p}_\nu ]=i \delta_{\mu,\nu}$.

Also, in this case, we derive the entanglement distance (ED) from a distance inspired by the Fubini-Study metric. For each normalized vector $|s\rangle \in {\cal H}^{cv}$, we have to consider the set of all the vectors derived from $|s\rangle$ under the action of LU operators.

In the case of CVS, dealing with infinite-dimensional Hilbert spaces and spanning the full set of LU operators is an impracticable task. Thus, the key point here is to determine, for each state $|s\rangle$, a set of LU operators that effectively generates variations suitable for the minimization of the trace. For instance, in the case of the class of states that are linear combinations of products of Glauber coherent states, an appropriate set of LU operators is the one of displacement operators. Thus, for each state $|s\rangle$ in this class, we consider
\begin{equation}
\left\lbrace |{\cal D},s\rangle = \prod^{M-1}_{\mu=0} {\cal D}^\mu |s\rangle \right\rbrace\, ,
\end{equation}
where each unitary operator ${\cal D}^\mu$ is a displacement operator
\begin{equation}
{\cal D}^\mu(\alpha^\mu) = \exp (\alpha^\mu a^\dagger_\mu - \alpha^{\mu *} a_\mu ) \, ,
\end{equation}
with $\alpha^\mu\in \mathbb{C}$. Also, we consider the differential operator given by
\begin{equation}
d{\cal D}^\mu := \sum^{n}_{\mu=1} d\tilde{\cal D}^\mu(d \xi^\mu) \, ,
\label{dDs}
\end{equation}
where for each differential operator $d\tilde{\cal D}^\mu$ results
\begin{equation}
d\tilde{\cal D}^\mu (d \xi^\mu)= d \xi^\mu a^\dagger_\mu - d \xi^{\mu *} a_\mu \, .
\end{equation}
We introduce the distance
\begin{equation}
\begin{split}
d^2_{FS}&=\langle {\cal D}, s | : d{\cal D}^{\mu \dagger}  d{\cal D}^\mu : | {\cal D}, s\rangle - |\langle {\cal D}, s |d{\cal D}^\mu  | {\cal D}, s\rangle|^2 \\
=&  \sum_{\mu \nu} (g_{\mu \nu} d\xi^{\mu *}  d\xi^\nu +
h_{\mu\nu} d\xi^{\mu }  d\xi^\nu  +
f_{\mu\nu} d\xi^{\mu *}  d\xi^{\nu*} 
)
\, ,
\end{split}
\end{equation}
where $::$ denotes the normal ordering operation. The entanglement distance (ED) for continuous variable systems (CVS) is therefore defined as
\begin{equation}
E(|s\rangle) =\inf_{ \{{\alpha}^\nu\}_\nu} \tr (g( |s\rangle , {\bm \alpha} )) \, ,
\label{emeasureCV}
\end{equation}
in analogy with \eqref{emeasure}. By direct calculations, one obtains
\begin{equation}
    E(|s\rangle) =4 \sum^{n}_{\mu =1}[
    \langle s | a^\dagger_\mu a_\mu |s\rangle -
     \langle s | a^\dagger_\mu |s \rangle \langle s| a_\mu |s\rangle]
    \, .
    \label{measureMqpureCV}
\end{equation}
Note that also in this case, $E(|s\rangle)$ has the meaning of a quantum correlation measure \cite{ScienticReports1-13}.

\section{Examples}
In this section, we apply our geometric method to investigate the entanglement properties of three classes of one/multi-parameter families of states.
In all the cases, the degree of entanglement of each state depends on these parameters and, in particular, they have known the values of the parameters corresponding to maximally entangled states for each one of the families.

The first, is a one-parameter family of states which is related to the  Greenberger-Horne-Zeilinger states \cite{ghz} since for a suitable choice of the parameter one gets a Greenberger-Horne-Zeilinger state.
We will name the elements of such family Greenberger-Horne-Zeilinger--like
states (GHZLS).
The second is a one-parameter family of states too. This class of states has been introduced by Briegel and Raussendorf in Ref. \cite{briegel_PRL86_910}, for this reason, we name the elements in this family Briegel-Raussendorf states (BRS).
The third is an $(M-1)$-parameters family of states,  related to the W states. In particular, we consider a weighted combination of the $M$ states that compose a W state of $M$-qubits. In this case, the state with the higher degree of entanglement is known to correspond to the case with the same weights.


In the following, we consider  the standard $M$-qubits 
basis
$\{
|0 \cdots  0 \rangle \, , \, \,
|0 \cdots 0 1 \rangle
, \ldots ,
|1 \cdots 1 \rangle \}
$ for ${\cal H}$, where $|0\rangle_\mu$ ($|1\rangle_\mu$) denotes the eigenstate of $\sigma^\mu_3$ with eigenvalue $+1$ ($-1$).
Thus, each basis' vector is identified by $M$ integers
$n_0,\ldots , n_{M-1} =0,1$ as
$
\ket{\{n\}} = |n_{M-1} \ n_{M-2} \quad n_0 \rangle \, 
$. Therefore, we enumerate such basis' vectors according to the binary integers representation
$ |k\rangle = \ket{\{n^k\}}$, with $k = \sum^{M-1}_{\mu=0} n^k_\mu 2^{\mu} $,
where $n^k_\nu$ is the $\nu$-th digit
of the number $k$ in binary representation and $k=0,\ldots,2^M-1$.

\subsection{Entanglement properties}

\subsubsection{Greenberger-Horne-Zeilinger--like states}
In this section, we consider a one-parameter family of states, the GHZLS,
defined according to
\begin{equation}
|GHZ,\theta \rangle_M = \cos(\theta) |0\rangle +\sin(\theta) 
|2^M-1\rangle \, ,
\label{ghz}
\end{equation}
where $0\leq\theta\leq \pi/2$.
For $\theta = 0,\pi/2$ the states are fully separable, whereas for
$\theta = \pi/4$ the state has the maximum degree of entanglement.
In this case, the trace for the metric tensor \eqref{gmunu} results
\begin{equation}
\begin{aligned}
\tr (g) &=M - \cos^2(2\theta)\sum^{M-1}_{\nu=0} 
(v^{\nu }_3)^2
  \, ,
 \end{aligned}
\label{g-ghzM}
\end{equation}
and, consistently with \eqref{vtildes}, it is minimised by the values $\tilde{v}^{\nu }_1 =\tilde{v}^{\nu }_2=0$, $\tilde{v}^{\nu }_3 =\pm 1$, for $\nu=0\,\ldots,M-1$. Therefore, we have
\begin{equation}
\tilde{g} = \sin^2(2\theta) J_M
\end{equation}
where $J_M$ is the $M\times M$ matrix of ones. The ED per qubit
for the GHZLS results
\begin{equation}
E(|GHZ,\theta\rangle_M)/M =\sin^2(2\theta) \, .
\label{ErGHZ}
\end{equation}

\subsubsection{Briegel Raussendorf states}
We denote with $\Pi^j_0=(\mathbb{I}+\sigma^j_3)/2$
and $\Pi^j_1=(\mathbb{I}-\sigma^j_3)/2$ the projector operators
onto the eigenstates of $\sigma^j_3$, $|0\rangle_j$ (with
eigenvalue $+1$)
and $|1\rangle_j$ (with eigenvalue $-1$),
respectively.
Each $M$ qubit state of the BRS class is
derived by applying to the fully separable state
\begin{equation}
|r, 0 \rangle = \bigotimes^{M-1}_{j=0}
\dfrac{1}{\sqrt{2}}(|0\rangle_j + |1\rangle_j)
  \, ,
\label{r0i}
\end{equation}
the non LU operator 
\begin{equation}
U_0(\phi) = \exp (-i \phi H_0) = \prod\limits^{M-1}_{j=1}
\left(
\mathbb{I} + \alpha \Pi^j_0 \Pi^{j+1}_1 
\right) \, ,
\label{U0e}
\end{equation}
where 
$
H_0 = \sum^{M-1}_{j=1}\Pi^j_0 \Pi^{j+1}_1 
$
and 
$
\alpha = (e^{-i\phi} -1) \, .
$
The full operator \eqref{U0e} is diagonal on the states of the standard
basis
$\{
|0 \cdots  0 \rangle \, , \, \,
|0 \cdots 0 1 \rangle
, \ldots ,
|1 \cdots 1 \rangle \}
$. 
In fact, the eigenvalue $\lambda_k$ of operator \eqref{U0e}, corresponding to a given
eigenstate $|k\rangle$ of this basis, results 
\begin{equation}
\lambda_k  = \sum^{n(k)}_{j=0} \binom{n(k)}{j} \alpha^j \, ,
\end{equation}
where $n(k)$ is the number ordered couples $01$ inside the sequence of the base
vector $|k\rangle$.
For the initial state \eqref{r0i} we consistently get 
\begin{equation}
|r, 0 \rangle_M =  2^{-M/2} \sum^{2^M-1}_{k=0} |k\rangle \, ,
\label{r0}
\end{equation}
and, under the action of $U_0(\phi)$ one
obtains
\begin{equation}
\begin{aligned}
|r, \phi \rangle_M = 
2^{-M/2} \sum^{2^M-1}_{k=0} 
\sum^{n(k)}_{j=0} \binom{n(k)}{j} \alpha^j
|k\rangle
 \, .
\label{state-phi}
\end{aligned}
\end{equation}
%
For $\phi =2 \pi k$, with $k\in \mathbb{Z}$, this state
is separable, whereas, for all the other choices of the
value $\phi$, it is entangled.
In particular, in \cite{briegel_PRL86_910} it is argued
that the values $\phi = (2k+1)  \pi$, where $k\in \mathbb{Z}$,
give maximally entangled states.

\vskip1.cm

\emph{Briegel Raussendorf states: case $M=2$}
\vskip0.5cm
In the case $M=2$ the one-parameter family of BRS is
\begin{equation}
|r, \phi \rangle_2 = \sum^3_{k=0 } c_k |k\rangle  \, ,
\label{rphi2}
\end{equation}
with $c_k = 1/2$ if $k\neq 1$, and $c_1 =  e^{-i\phi} /2$,
where $\phi \in [0, 2\pi]$.
By direct calculation one gets for the trace of the Fubini-Study metric
\begin{equation}
\tr (g) = \sum^1_{\nu=0}\left[1 - c^2\left(cv_1^\nu+\left(-1\right)^{\nu+1}sv_2^\nu\right)^2
 \right] \, ,
\label{g2}
\end{equation}
where $c= \cos\left({\phi}/{2}\right)$ and $s= \sin\left({\phi}/{2}\right)$.  Eq. \eqref{g2} is minimised with the choice $\tilde{\bf v}^\nu=\pm (c,(-1)^{\nu+1}s,0)$.
Consistently, the EM and the ED per qubit result 
\begin{equation}
\tilde{g} = s^2 J_2
\label{gtilde2}
\end{equation}
and 
\begin{equation}
E(|r, \phi \rangle_2)/2 = s^2\, .
\label{Erphi2}
\end{equation}
\vskip1.cm

\emph{Briegel Raussendorf states: case $M=3$}
\vskip0.5cm
In the case $M=3$ we have
\begin{equation}
|r, \phi \rangle_3 = \sum^7_{k=0} c_k |k\rangle \, ,
\label{rphi3}
\end{equation} 
with $c_k = 1/2^{3/2} $ if $k\neq 1,2,3,5$, and $c_k = e^{-i\phi} /2^{3/2}$ if 
$k= 1,2,3,5$,
where $\phi \in [0, 2\pi]$.
The trace of $g$ results
\begin{equation}
\tr (g) =\left[3 - \left( (c^2 v^0_1 + cs v^0_2)^2 + (c^2 v^1_1 )^2 + (c^2 v^2_1 - cs v^2_2)^2\right)
 \right] \, ,
\label{g3}
\end{equation}
is minimised with the choices $\tilde{\bf v}^0=(c,s,0)$, $\tilde{\bf v}^1=(1,0,0)$ and
$\tilde{\bf v}^2=(c,-s,0)$.
The EM and the ED per qubit, in this case result 
\begin{equation}
\tilde{g} ={s^2}
\left(
\begin{array}{ccc}
1 & c & 0\\
c & 1 +c^2 & c \\
0& c & 1
\end{array}
\right) \, ,
\label{gtilde3}
\end{equation}
and
\begin{equation}
E(|r, \phi \rangle_3)/3 = s^2 \left( 3+ c^2 \right)/3 \, ,
\label{Erphi3}
\end{equation}
respectively.

\vskip1.cm

\emph{Briegel Raussendorf states: case $M=4$}
\vskip0.5cm
For $M=4$ we have
\begin{equation}
|r, \phi \rangle_4 = \sum^{15}_{k=0} c_k |k\rangle \, ,
\end{equation}
with $c_k = e^{-i\phi} /4$ if $k\neq 0,5,8,12,14,15$, $c_5 = e^{-i2\phi} /4$, and
$c_0=c_8=c_{12}=c_{14}=c_{15}=1/4$,
where $\phi \in [0, 2\pi]$.
The trace of $g$ results
\begin{equation}
\begin{split}
\tr (g) =\left[4 - \left( (c^2 v^0_1 + cs v^0_2)^2 + (c^2 v^1_1 )^2 + \right. \right. \\
\left. \left. (c^2 v^2_1 )^2+ (c^2 v^3_1 - cs v^3_2)^2\right)
\right] \, ,
\end{split}
\label{g4}
\end{equation}
is minimised with the choices $\tilde{\bf v}^0=(c,s,0)$, $\tilde{\bf v}^1=(1,0,0)$, $\tilde{\bf v}^2=(1,0,0)$ and
$\tilde{\bf v}^3=(c,-s,0)$.
The EM in this case results
\begin{equation}
\tilde{g} ={s^2}
\left(
\begin{array}{cccc}
1 & c & 0 & 0\\
c & 1 +c^2 & c^2 & 0 \\
0& c^2 & 1 + c^2 & c \\
0 & 0 & c & 1
\end{array}
\right)\, ,
\label{gtilde4}
\end{equation}
thus the ED reads
\begin{equation}
E(|r, \phi \rangle_4)/4 = s^2 \left( 4+2 c^2 \right)/4 \,  .
\label{Erphi4}
\end{equation}

\subsubsection{$W$-states}
In this section, we consider an $(M-1)$-parameters family of states, the $W$ states,
defined according to
\begin{equation}
|W,{\bm \alpha} \rangle_M = \sum^M_{j=1} \alpha_j |2^{j-1}\rangle \, ,
\label{Wstate}
\end{equation}
with
\begin{equation}
\begin{split}
\alpha_1 = c_1 \\
\alpha_2 = s_1 c_2 \\
\vdots \\
\alpha_k = s_1 s_2 \cdots s_{k-1} c_k \\
\vdots \\
\alpha_{M-1} = s_1 s_2 \cdots s_{M-2} c_{M-1} \\
\alpha_{M} = s_1 s_2 \cdots s_{M-2} s_{M-1} \, ,
\end{split}
\end{equation}
where we have set $c_j=\cos(\theta_j)$, $s_j=\sin(\theta_j)$, and where $0\leq\theta_j\leq \pi/2$, $j=1,\ldots,M-1$.
If the number of indices $k$ such that $\alpha_k = 0$ is $r$, then the state \eqref{Wstate} is $r$-partite. For $\alpha_k = 1/\sqrt{M}$, for each $k$, the state \eqref{Wstate} is maximally entangled.

In the case of state \eqref{Wstate}, the trace of the metric tensor \eqref{gmunu} results
\begin{equation}
\begin{aligned}
\tr (g) &=\left[ M -\sum^{M-1}_{\nu=0} 
[(1-2|\alpha_{j(\nu)}|^2)v^{\nu }_3]^2
 \right] \, ,
 \end{aligned}
\label{g-ghzM}
\end{equation}
where $j(\nu)$ is an invertible map $j:\{0,\ldots,M-1\} \to \{1, \ldots, M \}$.
Consistently with \eqref{vtildes}, Eq. \eqref{g-ghzM} is minimised by the values $\tilde{v}^{\nu }_1 =\tilde{v}^{\nu }_2=0$, $\tilde{v}^{\nu }_3 =\pm 1$, for $\nu=0\,\ldots,M-1$. Therefore, the ED for these states results
\begin{equation}
E(|W,{\bm \alpha} \rangle_M)/M = 4 (1 -\sum_j |\alpha_j|^4)/M \, .
\label{ErW}
\end{equation}
The state with the higher entanglement corresponds to the choice $\alpha_j = 1/\sqrt{M}$, for $j=0,\ldots,M-1$. The corresponding value for the ED is
\begin{equation}
E(|W,{\bm \alpha_M} \rangle_M)/M = 4 (1 -1/M)/M \, .
\label{ErWmax}
\end{equation}
Therefore, exept for the case $M=2$, $E(|W,{\bm \alpha} \rangle_M)/M < 1$. This is due to the non-vanishing of the expectation values of the operators ${\bm \sigma}_{\bf v}^\mu$ on the state with the higher entanglement. In this sense, such state could not be considered a maximally entangled state. Remarkably, the $W$ state with maximal entanglement remains a maximal entanglement state under particle loss.

\vskip1.cm

\emph{$W$-states: case $M=2$}
\vskip0.5cm

For $M=2$ it is
\begin{equation}
|W,{\bm \alpha} \rangle_2 = \alpha_1 |1\rangle + \alpha_2 |2\rangle  \, ,
\end{equation}
where $\alpha_1 = \cos(\theta_1)$ and $\alpha_2 = \sin(\theta_1)$, and $\theta_1 \in [0,\pi/2]$.
The choice $\tilde{v}^{\nu }_1 =\tilde{v}^{\nu }_2=0$, $\tilde{v}^{\nu }_3 =(-1)^\nu$, for $\nu=0, 1$, by direct calculations one get the following expressions for the EM and ED
\begin{equation}
\tilde{g }= \sin^2(2\theta_1) J_2  \, ,
\label{EDW2}
\end{equation}
\begin{equation}
E(|W,{\bm \alpha} \rangle_2)/2 = \sin^2(2\theta_1)  \, ,
\label{EDW2}
\end{equation}
respectively.

\vskip1.cm

\emph{$W$-states: case $M=3$}
\vskip0.5cm
For $M=3$ it is
\begin{equation}
|W,{\bm \alpha} \rangle_3 = \alpha_1 |1\rangle + \alpha_2 |2\rangle  + \alpha_3
| 4\rangle\, ,
\end{equation}
where $\alpha_1 = \cos(\theta_1)$, $\alpha_2 = \sin(\theta_1) \cos(\theta_2)$ and $\alpha_3 = \sin(\theta_1) \sin(\theta_2)$.
By direct calculations one get the ED
\begin{equation}
\begin{split}
E(|W,{\bm \alpha} \rangle_3)/3 =& \dfrac{4}{3}  \left[ 1-
\left(  \cos^4(\theta_1) +  
\sin^4(\theta_1) \cos^4(\theta_2)  + 
\right. \right. \\ &\left. \left.
 \sin^4(\theta_1) \sin^4(\theta_2) 
\right)
\right]\, .
\label{EDW3}
\end{split}
\end{equation}

\subsection{Comparison between the families of states}

\subsubsection{Entanglement measure}
The ED $E(|r, \phi \rangle_M)/M$, as a function of $\phi$ and for $M=2,3,4$, gets the correct estimation of the degree of entanglement for the BRS one expect. In fact,
for fully separable states ($\phi=0,2\pi$) it vanishes, whereas for the
maximally entangled state, corresponding to $\phi=\pi$, the ED reaches the maximum (possible) value, $E(|r, \pi \rangle_M)/M=1$. This latter indicates that the expectation value  for any operator
$\tilde{\bf v}^\nu \cdot \boldsymbol{\sigma}^\nu $ ($\nu=0,\ldots,M-1$) vanishes on
the maximally entangled states of this class.

The ED for the GHZLS is null in the case of fully separable states
($\theta=0,\pi/2$) and gives the maximum value
 ($M$) in the case of the maximally entangled state ($\theta=\pi/4$). 
Also for the class of GHZLS, the expectation
value for the operators $\tilde{\bf v}^\nu \cdot \boldsymbol{\sigma}^\nu $ ($\nu=0,\ldots,M-1$) on the maximally entangled state is zero.

\begin{figure}[h]
\begin{center}
{ 
\includegraphics[width=1\linewidth]{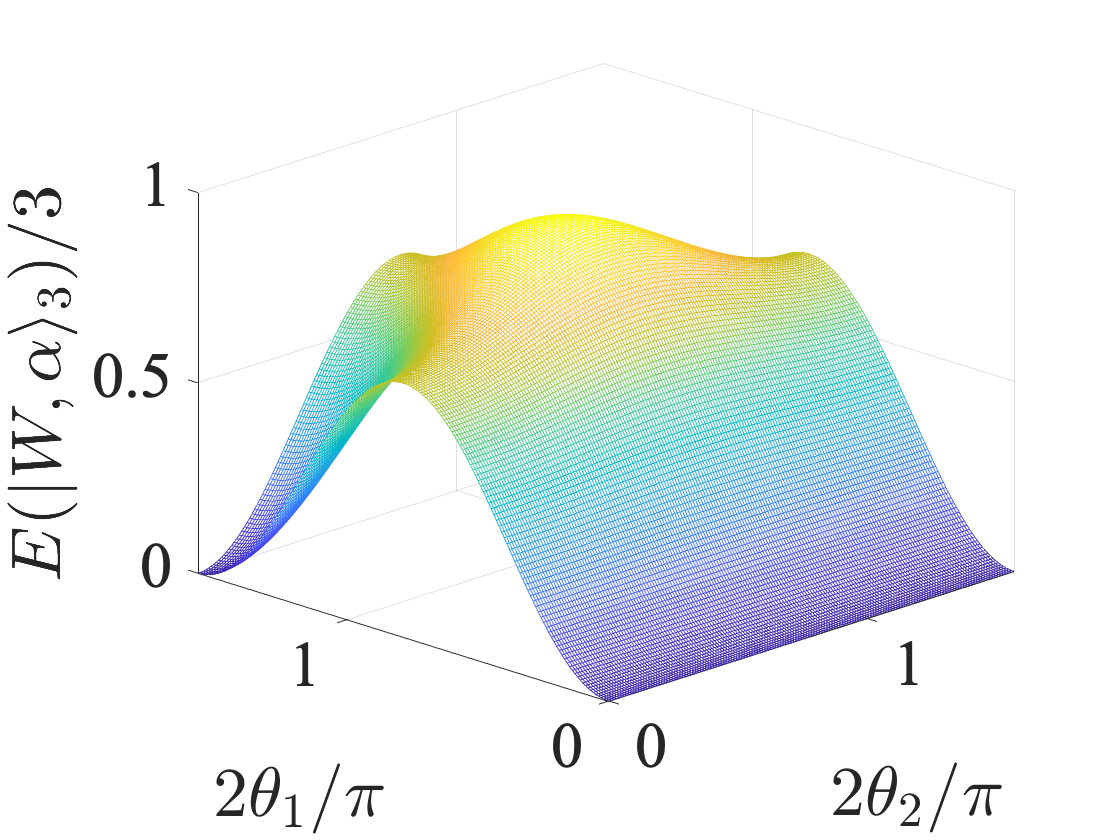}
}
\end{center}
\caption{This figure reports the three-dimensional plot of the ED $E(|W,{\bm \alpha} \rangle_3)/3$ as a function of $2 \theta_1/\pi$ and $2 \theta_2/pi$ for the states \eqref{Wstate}, in the case $M=3$.
}
\label{Fig05}
\end{figure}

In the case $M=2$, ${\bm \alpha}=(\cos(\theta_1),\sin(\theta_1))$, thus
for fully separable states ($\theta_1=0,\pi/2$) $E(|W,{\bm \alpha} \rangle_2)/2$  is null, whereas in the case $\theta_1=\pi/4$, it results
$E(|W,{\bm \alpha} \rangle_2)/2=1$.
In  Fig. \ref{Fig05} we consider the W-class of states for the case $M=3$. In this case
${\bm \alpha}=(\cos(\theta_1), \sin(\theta_1) \cos(\theta_2), \sin(\theta_1) \sin(\theta_2))$. In Fig. \ref{Fig05}  we report, with a $3D$ plot, the measure $E(|W,{\bm \alpha} \rangle_3)/3$ as a function of $\theta_1,\theta_2$ according to Eq. \eqref{ErW}, for the  states \eqref{Wstate}.
The measure \eqref{emeasure} catches in a surprisingly clear way the entanglement properties of this family of states. In particular, $E(|W,{\bm \alpha} \rangle_3)/3$
is null in the case of fully separable states ($\theta_1=0$).
The maximum entanglement $E(|W,{\bm \alpha} \rangle_3)/3=8/9$ is obtained in the case of the state given by $\theta_1 = \arccos(1/\sqrt{3})$ and $\theta_2= \arccos(1/\sqrt{2})$. $E(|W,{\bm \alpha} \rangle_3)/3 < 1$ indicates that the expectation value for the operators $\tilde{\bf v}^\nu \cdot \boldsymbol{\sigma}^\nu $ ($\nu=0,\ldots,M-1$) on the W states is different from zero. For this reason, the state with the maximum value of entanglement is not a maximally entangled state. Moreover, a lower value of the entanglement could be seen as a cons of this class of states, because the quantum correlation is less than in the case of the other states considered. Nevertheless, the W states are robust under the local measurements.
For the case of bi-separable states, ($\theta_2=0,\pi/2$) it results $0 <  E(|W,{\bm \alpha} \rangle_3)/3< 2/3$. 

\subsubsection{Equivalence classes characterization}
The present section is aimed to determine when a couple of states chosen into two different among the three families considered, actually belong to the same equivalence class. We compare the states with the highest entanglement of each family. 
Since the relevant geometric quantities for our analysis, in the case of GHZLS, are size-independent, we will assume the states of this family as the reference ones.
\vskip1.cm

\emph{Case $M=2$}
\vskip0.5cm
Let us start with the case $M=2$.
The maximally entangled state within the GHZLS is $|GHZ,\pi/4\rangle_2$, the unit vectors
that minimize the trace of the metric tensor are $\tilde{\bf v}_0=\tilde{\bf v}_1=(0,0,\pm 1)$, and the corresponding EM is $\tilde{g}(|GHZ,\pi/4\rangle_2)=J_2$.

In the case of the second family, the maximally-entangled state is $|r,\pi\rangle_2$, the unit vectors that minimize the trace of $g$ are $\tilde{\bf v}^0=\pm(0,-1,0)$ and $\tilde{\bf v}^1=\pm(0,1,0)$. The corresponding EM results $\tilde{g}(|r,\pi\rangle_2)=J_2 = \tilde{g}(|GHZ,\pi/4\rangle_2)$. Therefore, according to our criterion, $|GHZ,\pi/4\rangle_2$ and $|r,\pi\rangle_2$ belong to the same equivalence class:
$ [|GHZ,\pi/4\rangle_2] = [|r,\pi\rangle_2]$. One can verify the correctness of this result since by direct calculation one gets $|GHZ,\pi/4\rangle_2=e^{-i \sigma^1_2\pi/4}|r,\pi\rangle_2$.

For the third family, the maximally-entangled state is $|W,(1/\sqrt{2},1/\sqrt{2})\rangle_2$, the unit vectors that minimize the trace of $g$ are $\tilde{\bf v}^0=\pm(0,0,1)$ and $\tilde{\bf v}^1=\pm(0,0,-1)$. The corresponding EM results $\tilde{g}(|W,(1/\sqrt{2},1/\sqrt{2})\rangle_2)=J_2 = \tilde{g}(|GHZ,\pi/4\rangle_2)$. Therefore, according to our criterion, $[|GHZ,\pi/4\rangle_2]=[|W,(1/\sqrt{2},1/\sqrt{2})\rangle_2]$. One can verify the correctness of this result since by direct calculation one gets $|GHZ,\pi/4\rangle_2=e^{-i\pi/2}e^{-i \sigma^0_1\pi/4}e^{-i \sigma^1_1\pi/4}|W,(1/\sqrt{2},1/\sqrt{2})\rangle_2$.
\begin{figure}[h]
\begin{center}
{ 
\includegraphics[width=1\linewidth]{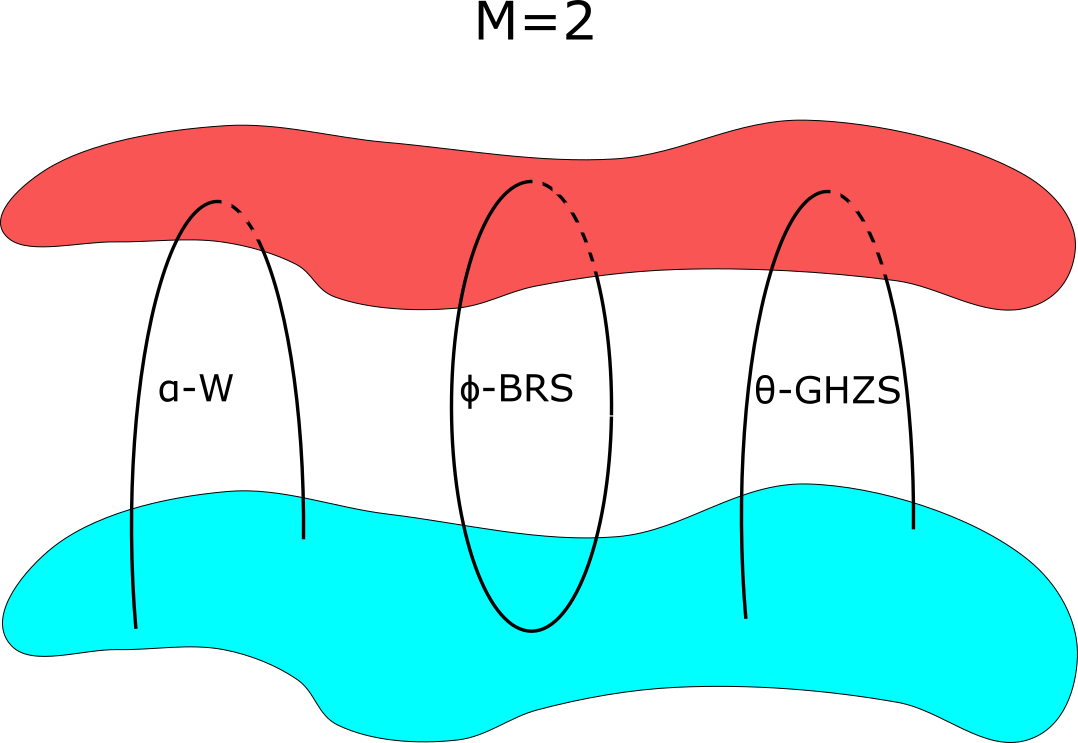}
}
\end{center}
\caption{The scheme in the figure represents the topological structure of the equivalence classes for some of the states for each of the three families. A point along one of the black lines represents a state of each family, from left to right these are the $\bm{\alpha}-W$ states, the $\phi$-BRS and the $\theta$-GHZLS. The magenta cloudlet represents the equivalence class to which belong the (fully separable) states. The latter are, in the case of the $\bm{\alpha}$-W states, the states $|W,\bm{\alpha}\rangle_2$ with $\bm{\alpha} = (1,0), (0,1)$, in the case of $\phi$-BRS, the states $|r,\phi\rangle_2$ with $\phi =0,2\pi$ and, in the case of the $\theta$-GBZLS the states $|GHZ,\theta\rangle_2$, where $\theta=0,\pi/2$. In the case of $M=2$ qubits, the states of the three families with the higher degree of entanglement, that is $|W,(1/\sqrt{2},1/\sqrt{2})\rangle_2$, $|r,\pi\rangle_2$ and $|GHZ,\pi/4\rangle_2$, belong to the same equivalence class, figured with a red cloudlet.
}
\label{M2}
\end{figure}

\vskip1.cm

\emph{Case $M=3$}
\vskip0.5cm

Let us consider the case $M=3$.
The maximally entangled state within the GHZLS is $|GHZ,\pi/4\rangle_3$, the unit vectors
that minimize the trace of the metric tensor are $\tilde{\bf v}_0=\tilde{\bf v}_1=\tilde{\bf v}_2=(0,0,\pm 1)$, and the corresponding EM is $\tilde{g}(|GHZ,\pi/4\rangle_3)=J_3$.

In the case of the second family, the maximally-entangled state is $|r,\pi\rangle_3$.
In this case $\tilde{g} (|r,\pi\rangle_3)\neq \tilde{g}(|GHZ,\pi/4\rangle_3)$, as  shown in Eq. \eqref{gtilde3}.
Nevertheless, we have
\begin{equation}
g(|r,\pi\rangle_3,{\bf v}) = 
\left(
\begin{array}{ccc}
1 						& v^0_1 v^1_3 &  -v^0_1 v^2_1  \\
 v^0_1 v^1_3 & 1 					&  -v^1_2 v^2_1 \\
-v^0_1 v^2_1 &   -v^1_3 v^2_1& 1
\end{array}
\right) \, ,
\end{equation}
thus, the choice for the unit vectors ${\bf v}^0=\pm(1,0,0)$, ${\bf v}^1=\pm(0,0,1)$ and ${\bf v}^2=\mp(1,0,0)$ gives $g(|r,\pi\rangle_3,{\bf v}) = J_3$.
Therefore, according to our criterion, $|GHZ,\pi/4\rangle_3$ and $|r,\pi\rangle_3$ might belong to the same equivalence class.
In fact by direct calculation one can verify that $|GHZ,\pi/4\rangle_3=e^{i \sigma^0_2\pi/4}e^{-i \sigma^2_2\pi/4}|r,\pi\rangle_3$, hence $[|GHZ,\pi/4\rangle_3]=[|r,\pi\rangle_3]$.

\begin{figure}[h]
\begin{center}
{ 
\includegraphics[width=1\linewidth]{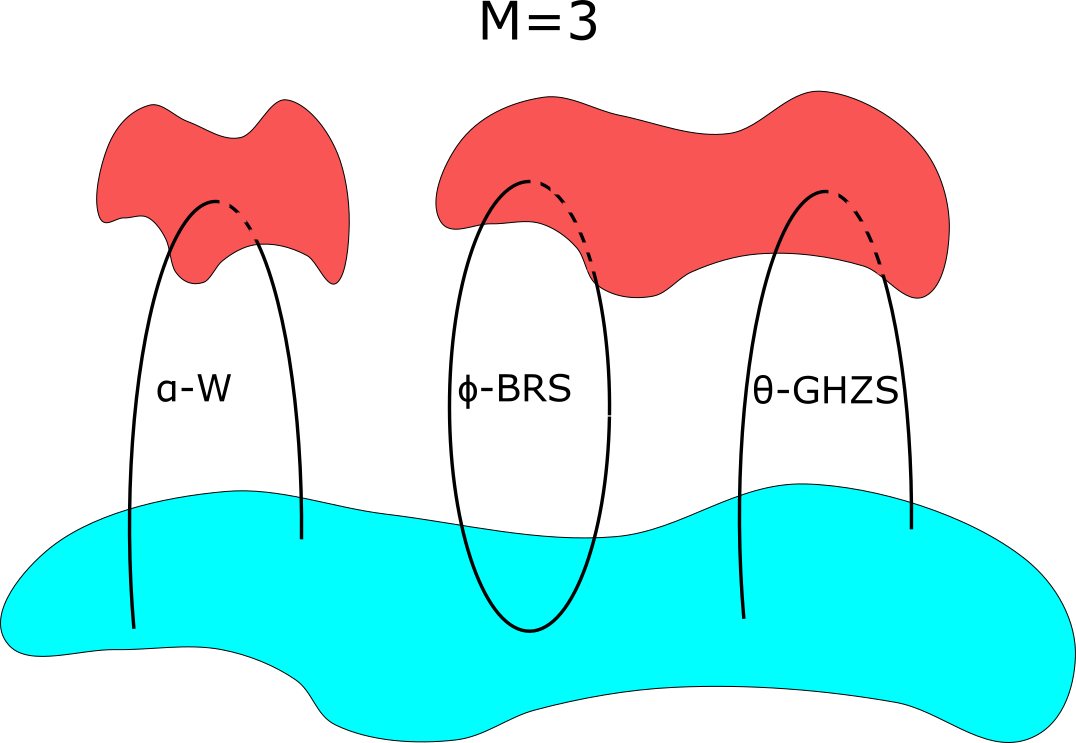}
}
\end{center}
\caption{As in Fig. \ref{M2}, we report here a scheme that represents the topological structure of the equivalence classes for some of the states of the three families: the $\bm{\alpha}-W$ states, the $\phi$-BRS and the $\theta$-GHZLS. In this case, we consider three qubits states. The magenta cloudlet represents the equivalence class to which belong the (fully separable) states for the three families:
$|W,\bm{\alpha}\rangle_3$ with $\bm{\alpha} = (1,0,0), (0,1,0),(0,0,1)$, $|r,\phi\rangle_3$ with $\phi =0,2\pi$ and, $|GHZ,\theta\rangle_3$, with $\theta=0,\pi/2$. In the case of $M=3$ qubits, the equivalence classes of the states of the three families with the higher degree of entanglement do not coincide. In fact, the class $[|W,(1/\sqrt{3},1/\sqrt{3},1/\sqrt{3})\rangle_3]$ is dijointed from the class $[|r,\pi\rangle_3]=[|GHZ,\pi/4\rangle_3]$, as figured with the two red cloudlets.
}
\label{M3}
\end{figure}

For the third family, the state with the higher entanglement is $|W,(1/\sqrt{3},1/\sqrt{3},1/\sqrt{3})\rangle_3$. The latter is not a maximally entangled state since $E(|W,(1/\sqrt{3},1/\sqrt{3},1/\sqrt{3})\rangle_3)/3=8/9< 1$.
The unit vectors that minimize the trace of $g$ are $\tilde{\bf v}^\nu=\pm(0,0,1)$ for $\nu=0,1,2$. The corresponding EM results 
\begin{equation}
\tilde{g}(|W,(1/\sqrt{3},1/\sqrt{3},1/\sqrt{3})\rangle_3) = 
\left(
\begin{array}{ccc}
\frac{2}{3} & - \frac{4}{9} & -\frac{4}{9} \\
-\frac{4}{9} & \frac{2}{3} & -\frac{4}{9} \\
-\frac{4}{9} & -\frac{4}{9} & \frac{2}{3}
\end{array}
\right) \, .
\end{equation}
Furthermore, no choice of the unit vectors leads to the expression $J_3$ for the metric tensor. Therefore, upon our criterion, the states of the third family and the ones of the first two families are inequivalent: $[|W,(1/\sqrt{3},1/\sqrt{3},1/\sqrt{3})\rangle_3]\neq [|GHZ,\pi/4\rangle_3]$. This result agrees with the study of Ref. \cite{PhysRevA.62.062314}.

\vskip1.cm

\emph{Case $M=4$}
\vskip0.5cm
As last we consider the case $M=4$.
The maximally entangled state within the GHZLS is $|GHZ,\pi/4\rangle_4$, the unit vectors
that minimize the trace of the metric tensor are $\tilde{\bf v}_0=\tilde{\bf v}_1=\tilde{\bf v}_2=\tilde{\bf v}_3=(0,0,\pm 1)$, and the corresponding EM is $\tilde{g}(|GHZ,\pi/4\rangle_4)=J_4$.
In the case of the second family, the maximally-entangled state is $|r,\pi\rangle_4$.
This is a genuine maximally entangled state since $E(|r, \pi \rangle_4)/4 = 1$.
We have $\tilde{g}(|r,\pi\rangle_4) \neq \tilde{g}(|GHZ,\pi/4\rangle_4)$, as  shown in Eq. \eqref{gtilde4}.
We have
\begin{equation}
g(|r,\pi\rangle_4,{\bf v}) = 
\left(
\begin{array}{cccc}
1 						& v^0_1 v^1_3 &  0 & 0 \\
 v^0_1 v^1_3 & 1 					&  0 & 0 \\
0 &   0& 1 &  - v^2_3 v^3_1\\
0&0&- v^2_3 v^3_1& 1
\end{array}
\right) \, ,
\end{equation}
thus, in this case, no choice of the unit vectors leads to an expression for the metric tensor equivalent to $J_4$. 
Therefore, we conclude $[|GHZ,\pi/4\rangle_4] \neq [|r,\pi\rangle_4]$. This confirms the result reported in Ref. \cite{PhysRevA.63.012308}.

\begin{figure}[h]
\begin{center}
{ 
\includegraphics[width=1\linewidth]{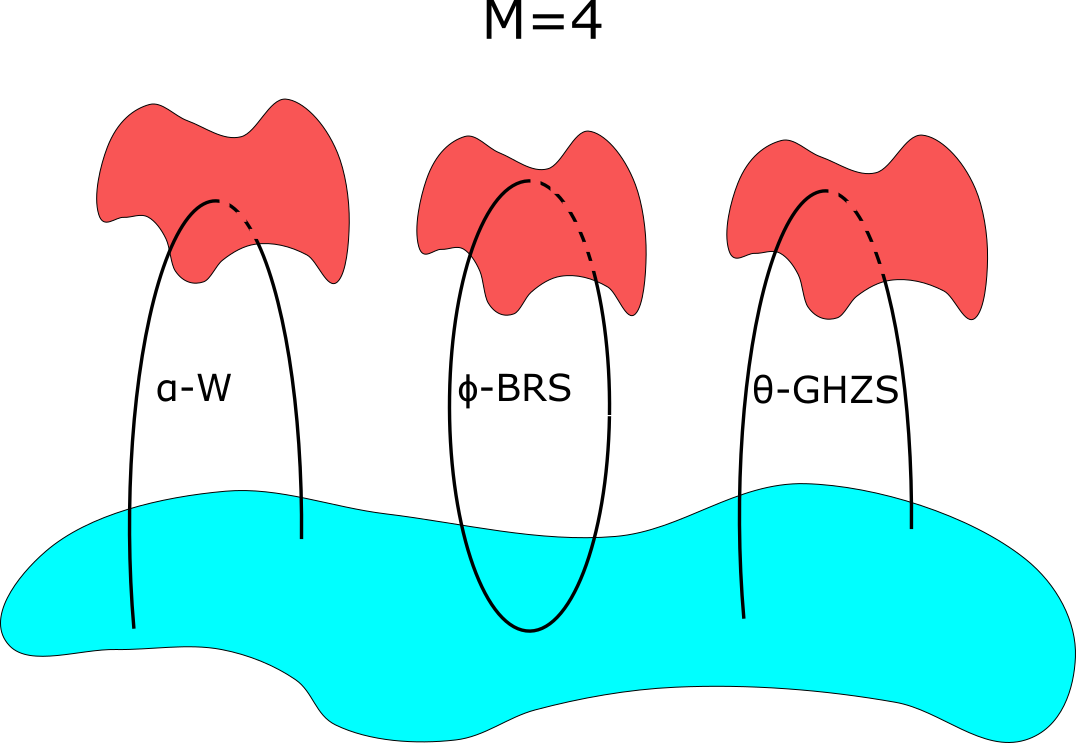}
}
\end{center}
\caption{We report here a scheme analogous to those of Figs. \ref{M2} and \ref{M3}. In this case, we consider $M=4$ qubits states. The magenta cloudlet represents the equivalence class to which belong the (fully separable) states of the three families: 
$|W,\bm{\alpha}\rangle_4$ with $\bm{\alpha} = (1,0,0,0), (0,1,0,0),(0,0,1,0), (0,0,0,1)$, $|r,\phi\rangle_4$ with $\phi =0,2\pi$ and, $|GHZ,\theta\rangle_4$, with $\theta=0,\pi/2$. In this case, the equivalence classes of the states of the three families with the higher degree of entanglement are all disjointed. In fact, the class $[|W,(1/\sqrt{4},1/\sqrt{4},1/\sqrt{4},1/\sqrt{4})\rangle_4]\neq [|r,\pi\rangle_4]\neq [|GHZ,\pi/4\rangle_4]$, as figured with the three dijointed red cloudlets.
}
\label{M4}
\end{figure}

\section{Glauber coherent  states}
Let us consider the state
\begin{equation}
|s\rangle=c  (|\alpha_1, \alpha_2\rangle + |\alpha_2, \alpha_1\rangle) \, ,
\end{equation}
where $\alpha_1,\alpha_2 \in \mathbb{C}$,  $|\alpha_j, \alpha_k\rangle = |\alpha_j\rangle |\alpha_k\rangle $ is product
of Glauber coherent states 
\begin{equation}
|\alpha\rangle = e^{-|\alpha|^2/2} \sum^\infty_{n=0} \dfrac{\alpha^n}{\sqrt{n !}} |n\rangle  \, ,
\end{equation}
and $c$ is the  normalization factor.
We define
\begin{equation}
p =\langle \alpha_1, \alpha_2|\alpha_2, \alpha_1\rangle 
= e^{-|\alpha_1- \alpha_2|^2 }
\, .
\end{equation}
The state $|s\rangle$ is separable if and only if $\alpha_1=\alpha_2$, that is $p=1$.
By direct calculations we got from Eq. \eqref{measureMqpureCV} the following expression for the ED
\begin{equation}
E(|s\rangle) = 2\dfrac{1-p}{1+p} 
|\alpha_1- \alpha_2|^2
\, ,
\end{equation}
that agrees to the prediction, of being null if and only if $\alpha_1=\alpha_2$.

\section{Concluding remarks}

In the present work, we have investigated the deep link between the Riemannian metric structure associated with the projective Hilbert space and the entanglement measure for the states within this space. We first considered the case of a general multi-qubit system and then examined the case of a continuous variables system represented by the product of two Glauber coherent states. In particular, we have shown that entanglement has a remarkable geometrical interpretation. In fact,  we have shown that the ED of a state $|\psi\rangle$ is the minimum of the sum of the squared distances between $|\psi\rangle$ and all its conjugate states ${\bf v}^\mu \cdot {\bm \sigma}^\mu |\psi\rangle$, where ${\bf v}^\mu$ are unit vectors and $\mu$ runs on the number of parties. In such a sense, entanglement is an obstacle to the minimum of the sum of the distances between $|\psi\rangle$ and its conjugate states.

Also, within the proposed geometric approach, we have derived a general method to determine if two states are not LU-equivalent. For bipartite states, it further allows to confirm if they are, on the contrary, LU-equivalent.

The entanglement measure named ED, proposed in Ref. \cite{PhysRevA.101.042129}, has the desirable property of providing a directly computable measure of entanglement for a general multi-qudit hybrid pure state. A convex roof extension of it to the most general case of mixed states can easily be built.  
In the present work, we have proved that the ED of a state $|\psi\rangle$ {\it i)} is null if and only $|\psi\rangle$ is fully separable; {\it ii)} is invariant under LU operations; {\it iii)} doesn't increase, in average, under LOCC. This, definitely validate the entanglement distance as an appropriate entanglement measure for multipartite states, pure or mixed.

Finally, we have applied the proposed geometric approach to the study of the entanglement magnitude and the equivalence classes properties, of three families of pure states. 

\begin{acknowledgments}
We acknowledge support from the RESEARCH SUPPORT PLAN 2022 - Call for applications for funding allocation to research projects curiosity driven (F CUR) - Project ”Entanglement Protection of Qubits’ Dynamics in a Cavity”– EPQDC and the support by the Italian National Group of Mathematical Physics (GNFM-INdAM). 
\end{acknowledgments}


\appendix

\section*{Appendix}
\subsection*{Monotonicity, on average, under unilocal quantum operation}

In Ref. \cite{vidal_2000}, it is shown that LOCC may be decomposed as series of unilocal quantum operations (UQ). It results that it is enough, to prove condition \ref{tre}, to show that the measure $E(\rho)$ \begin{itemize}\item is non-increasing, on average, under UQ \item is convex as a measure on mixed states. \end{itemize}
The second condition is automatically fullfilled by the convex roof construction \eqref{mixedstate_ED}.
The first one can be decomposed into the four following conditions:
\begin{enumerate}[label=(\alph*)]
\item $E$ is LU-invariant, that is $E(\rho)=E(U\rho U^\dagger)$.\label{UQ:LU}
\item $E$ is non-increasing on average under (non necessarily complete) unilocal Von Neumann measurement, that is $E(\rho)\geq \sum_j p_j E(\rho_j)$, with $\rho_j$ is one of the outcomes of such a measurement, with associated probability $p_j$. \label{UQ:VNmeas}
\item $E$ is invariant under the addition of an uncorrelated ancilla $A$, that is $E(\rho)=E(\rho\otimes\rho_A)$, with $\rho_A$ the state of $A$.\label{UQ:ancilla-add}
\item $E$ is non-increasing under the removal of any local part $A$ of the system, that is $E(\rho)\geq E(\Tr_A(\rho))$. \label{UQ:ancilla-remov}
\end{enumerate}

Condition \ref{UQ:LU} holds by construction, as shown in Eq. \eqref{LUinv_proof}.

Unilocal Von Neumann measurements can be realized by completely positive map $\Theta$, that converts $|\psi\rangle$ to $M_j|\psi\rangle$ with probability $p_j =\langle \psi |M^\dagger_j M_j | \psi\rangle$,  where $\sum_j M^\dagger_j M_j \leq\mathbb{I}$. Therefore, condition \ref{tre} reads
\begin{equation}
E(|\psi\rangle) \geq \sum_j p_j E(|\psi_j\rangle) \, ,
\end{equation}
where $|\psi_j\rangle = M_j|\psi\rangle/\sqrt{p_j}$.
The proof of this inequality follows from the following calculation. To make easier the formulas, in the following we will indicate the expectation value of an operator ${\cal O}$ on a state $|\psi\rangle$ with $\langle {\cal O}\rangle_\psi$.
Let $\bar{\mu}$ be the qubit on which the generalized measurement $\{M_j\}_j$ operates.
Let us drop the index $\bar{\mu}$ for the Kraus operators $\{M_j,M^\dagger_j\}_j$ for ease of notations.
In the case of a general state $|\psi\rangle$ and for $\mu\neq\bar{\mu}$ it results $[M_j,\sigma^\mu_k]=0$ for any $j$ and $k$. Consistently, from Eq. \eqref{gmunu} we have
\begin{equation}
\begin{aligned}
&g_{\mu \mu}  (|\psi\rangle,{\bf v}^\mu) = \langle \left( \bm{\sigma}_{\bf v}^\mu -
 \langle  \bm{\sigma}_{\bf v}^\mu \rangle_\psi \right )^2 \rangle_\psi \\
& \geq \langle \sum_j M^\dagger_j M_j\left( \bm{\sigma}_{\bf v}^\mu -
 \langle \bm{\sigma}_{\bf v}^\mu \rangle_\psi \right )^2\rangle_\psi \\
& = \sum_j   \langle M^\dagger_j \left( \bm{\sigma}_{\bf v}^\mu -
 \langle  \bm{\sigma}_{\bf v}^\mu\rangle_\psi \right )^2 M_j\rangle_\psi \\
& = \sum_j p_j  \langle \left( \bm{\sigma}_{\bf v}^\mu -
 \langle  \bm{\sigma}_{\bf v}^\mu\rangle_\psi \right )^2 \rangle_{\psi_j} \\
& = \sum_j  p_j  \langle  \left( \bm{\sigma}_{\bf v}^\mu \!-\!
\langle  \bm{\sigma}_{\bf v}^\mu\rangle_{\psi_j} \!+\!
\langle  \bm{\sigma}_{\bf v}^\mu\rangle_{\psi_j} \!-\!
 \langle  \bm{\sigma}_{\bf v}^\mu\rangle_\psi \right )^2\rangle_{\psi_j} \\
 & =\sum_j p_j \left[ \langle  \left( \bm{\sigma}_{\bf v}^\mu -   \langle \bm{\sigma}_{\bf v}^\mu \rangle_{\psi_j}  \right )^2\rangle_{\psi_j}\right. + \left. \langle \left( 
\langle  \bm{\sigma}_{\bf v}^\mu\rangle_{\psi_j} -
 \langle  \bm{\sigma}_{\bf v}^\mu\rangle_\psi 
 \right )^2 \rangle_{\psi_j} \right]
  \\
 &\geq
  \sum_j p_j \langle  \left( \bm{\sigma}_{\bf v}^\mu -   \langle \bm{\sigma}_{\bf v}^\mu \rangle_{\psi_j}  \right )^2\rangle_{\psi_j}  =
   \sum_j p_j g_{\mu \mu}  (|\psi_j\rangle,{\bf v}^\mu) \, .
\end{aligned}
\end{equation}
In summary, for $\mu\neq\bar{\mu}$ it results
\begin{equation}
g_{\mu \mu}  (|\psi\rangle,{\bf v}^\mu) \geq  \sum_j p_j g_{\mu \mu}  (|\psi_j\rangle,{\bf v}^\mu) 
\end{equation}
and then
\begin{equation}
\begin{aligned}
\inf_{ 
{\bf v}^\mu} g_{\mu \mu}  (|\psi\rangle,{\bf v}^\mu) & \geq \inf_{ 
{\bf v}^\mu}   \sum_j p_j g_{\mu \mu}  (|\psi_j\rangle,{\bf v}^\mu) \geq \\
& \sum_j p_j  \inf_{ 
{\bf v}^\mu_j} g_{\mu \mu}  (|\psi_j\rangle,{\bf v}^\mu_j) \, .
\end{aligned}
\end{equation}
We note that in the case general case it is
\begin{equation}
0 \leq g_{\mu \mu}  (|\psi\rangle,{\bf v}) \leq 1 \, .
\end{equation}

Let us now consider the contribution to the entanglement of the state of a qubit $\mu=\bar{\mu}$.
In this case we have
\begin{equation}
 \sum_{j} p_j \langle    \left( \bm{\sigma}_{\bf v}^{\bar{\mu}} -
 \langle \bm{\sigma}_{\bf v}^{\bar{\mu}} \rangle_{\psi_j} \right )^2 \rangle_{\psi_j}  \, ,
\end{equation}
in which, for each possible outcome of the measurement $j$, it is possible to choose a unit vector ${\bf v}^{\bar{\mu}}={\bf v}^{\bar{\mu}}_j$ so that 
\begin{equation}
\bm{\sigma}_{{\bf v}_j}^{\bar{\mu}} |\psi_j\rangle = |\psi_j\rangle \, .
\end{equation}
Therefore it results
\begin{equation}
\inf_{ 
{\bf v}^{\bar{\mu}}}  g_{\bar{\mu} \bar{\mu}}(|\psi_j\rangle,{\bf v}^{\bar{\mu}}) = 0 \, .
\end{equation}
This proves that the contribution to the entanglement of the state $|\phi_j\rangle$ coming from the $\bar{\mu}$th qubit is null.

Finally, we have then
\begin{equation}
\begin{aligned}
E(|\psi\rangle) = & \inf_{ 
\{
{\bf v}^\mu\}_\mu} \sum^{M-1}_{\mu=0} g_{\mu \mu}  (|\psi\rangle,{\bf v}^\mu) \geq \\
&\sum_j p_j \sum^{M-1}_{\mu=0}  \inf_{ 
{\bf v}^\mu_j} g_{\mu \mu}  (|\psi_j\rangle,{\bf v}^\mu_j) =\sum_j p_j E(|\psi_j\rangle)  \, ,
\end{aligned}
\end{equation}
that proves condition \ref{UQ:VNmeas}.\\

Condition \ref{UQ:ancilla-add} is clearly satisfied for pure states, and this property is preserved by the convex roof construction \eqref{mixedstate_ED}.\\

Finally, note that the reduced density matrix for qubit $\mu$ writes 
\begin{equation}
\rho^\mu=\Tr_{\mu^c}\big[|\psi\rangle\langle\psi|\big]=\frac{1}{2}(\mathbb{I}+\tilde{\bf v}^\mu\cdot\bm{\sigma}^\mu) ,
\end{equation} where $\mu^c$ is the complement of $\mu$ on the set of all qubits in the system, and $\tilde{\bf v}^\mu=\langle\psi|\bm{\sigma}^\mu|\psi\rangle$. Also remark that $\Tr\big[(\rho^\mu)^2\big]=\frac{1}{2}(1 + |\tilde{\bf v}^\mu|^2)$.

Yet, from Eq. \eqref{measureMqpure}, the single-qubit measure follows
\begin{equation}\label{single-q-ED_part-trace}
\begin{split}
E_\mu(|\psi\rangle)&=\min\limits_{{\bf v}^\mu}g_{\mu\mu}(|\psi\rangle,{\bf v}^\mu)\\
&=1-|\langle\psi|\bm{\sigma}^\mu|\psi\rangle|^2=1-|\tilde{\bf v}^\mu|^2\\
&=2\Bigg(1-\Tr\Big[\Big(\Tr_{\mu^c}\big[|\psi\rangle\langle\psi|\big]\Big)^2\Big]\Bigg),
\end{split}
\end{equation}
which is invariant with respect to the partial trace applied to any subsystem $A\in\mu^c$, since this information is discarded anyway, and we have $E\Big(|\psi\rangle\langle\psi|\Big)= E\Big(\Tr_A\big[|\psi\rangle\langle\psi|\big]\Big)$. Condition \ref{UQ:ancilla-remov} is hence fulfilled by the pure state ED.

After such a removal, it can easily be checked that the convex roof construction \eqref{mixedstate_ED} rewrites
\begin{equation}
E_\mu\Big(\Tr_A\big[\rho\big]\Big)=\min_{\{p_j,\psi_j\}}\sum_j p_j E_\mu\Big(\Tr_A\big[|\psi_j\rangle\langle\psi_j|\big]\Big)
\end{equation}
for any $\rho=\sum_j p_j |\psi_j\rangle\langle\psi_j|$. Thus, $E_\mu\Big(\Tr_A\big[\rho\big]\Big)= E_\mu(\rho)$, so $E(\rho)\geq E\Big(\Tr_A\big[\rho\big]\Big)$, proving that condition \ref{UQ:ancilla-remov} also holds for the mixed states measure.

\subsection*{Convexity of the function of the partial trace}

In Ref. \cite{vidal_2000}, another necessary and sufficient condition is proposed for a magnitude to be LOCC-monotone on average. Namely, any function $E_\mu(\psi)=f(\Tr_\mu(\psi))$ such that $f$ is concave and LU-invariant, is an entanglement monotone for pure states. In addition, the related mixed state measure obtained by convex roof construction is a LOCC-monotone on average, i.e. an entanglement monotone.

As noticed in Eq. \eqref{single-q-ED_part-trace}, $E_\mu(|\psi\rangle)=f(\Tr_{\mu^c}[\psi])$ with $f(\bm{x})= 2(1 - \Tr[\bm{x}^2])$. 

$f$ is clearly LU-invariant. Let us now prove that $f$ is concave. Consider $\rho_k=\frac{1}{2}(\mathbb{I}+{\bf n}_k\cdot\bm{\sigma})$, with $k=1,2$, two single-qubit density matrices, and $\lambda\in[0,1]$. We have
\begin{equation}
\begin{split}
f\Big(\lambda\rho_1+(1-\lambda)\rho_2\Big)
=1 - |\lambda{\bf n}_1 + (1-\lambda){\bf n}_2|^2
\end{split}
\end{equation} 
and
\begin{equation}
\lambda f(\rho_1)+(1-\lambda)f(\rho_2)=1 - \lambda|{\bf n}_1|^2 + (1-\lambda)|{\bf n}_2|^2,
\end{equation} 
and the convexity of the Euclidean squared norm implies
\begin{equation}
f\Big(\lambda\rho_1+(1-\lambda)\rho_2\Big)\geq\lambda f(\rho_1)+(1-\lambda)f(\rho_2),
\end{equation}
that is, $f$ is concave, which completes our proof that $E_\mu(|\psi\rangle)$ is a valid entanglement monotone for pure states. \\
According to Ref. \cite{vidal_2000}, Eq. \eqref{mixedstate_ED} is then itself an entanglement monotone.

\hfill
\bibliography{references}

\end{document}